# Magnetic properties and atomic structure of $La_{2/3}Ca_{1/3}MnO_3$-$YBa_2Cu_3O_7$ heterointerfaces

Z. L. Zhang[1#], U. Kaiser [1], S. Soltan [2,3], H-U. Habermeier [2], B. Keimer [2]

1. Electron Microscopy Group for Materials Science, University of Ulm, D-89091 Ulm, Germany
2. Max Planck Institute for Solid State Research, D-70561 Stuttgart, Germany
3. Physics Department, Faculty of Science, Helwan University, 11795-Cairo, Egypt

## Abstract

A heterostructure comprised of a 2.7 nm (7 unit-cell) thick layer of the metallic ferromagnet $La_{2/3}Ca_{1/3}MnO_3$ and two 50 nm thick layers of the high-temperature superconductor $YBa_2Cu_3O_7$ epitaxially grown on (100) $SrTiO_3$ by pulsed-laser deposition was characterized by magnetization measurements and spherical-aberration-corrected high-resolution transmission electron microscopy (HRTEM). The saturation magnetization is about half of that in bulk $La_{2/3}Ca_{1/3}MnO_3$. A massive reduction of the magnetization previously inferred from sputter-deposited $La_{2/3}Ca_{1/3}MnO_3$-$YBa_2Cu_3O_7$ heterostructures can be ruled out. HRTEM image analysis, combined with image simulation and a focus series reconstruction, revealed atomically sharp epitaxial structures with stacking sequences -(La,Ca)O-$CuO_2$- and -BaO-MnO- at the top and bottom interface.

# Current address: Erich Schmid Institute of Materials Science, Austrian Academy of Sciences, Jahnstraße 12, A-8700 Leoben, Austria

Interfaces between ferromagnets and superconductors are of topical interest because of unusual proximity effects that may find device applications in superconducting electronics [1]. Recent advances in the synthesis of epitaxial transition metal oxide heterostructures offer intriguing perspectives for the practical realization of such interfaces. In particular, high-quality heterostructures of the metallic ferromagnet $La_{2/3}Ca_{1/3}MnO_3$ (LCMO) and the high-temperature superconductor $YBa_2Cu_3O_7$ (YBCO) have been prepared by pulsed-laser deposition (PLD) and investigated with a variety of experimental methods [2-15]. While interesting macroscopic properties such as exchange bias [7] and giant magnetoresistance [10] were observed in some of these structures, the presence of microscopic proximity effects is still under intense investigation [3]. Some experiments have yielded evidence of a strongly suppressed ferromagnetic magnetization on the LCMO side of the interface [5,12], whereas others conclude that copper spins on the YBCO side of the interface are polarized by exchange interactions across the interface, partially compensating the ferromagnetic magnetization of LCMO [13]. Interplay between ferromagnetic and superconducting order parameters [1] is unlikely to be responsible for either of these effects, because the microscopic magnetic properties are not noticeably affected by the superconducting transition [15]. Rather, charge transfer across the interface [12,14] and an orbital reconstruction due to the formation of a covalent bond across the interface [14] have been invoked as explanations.

The atomic stacking sequence at the interface is an important ingredient in microscopic models of these proximity effects and other physical properties of oxide heterostructures. Early scanning transmission electron microscopy (STEM) experiments [6] had revealed two stacking sequences at the YBCO-LCMO interface: YBCO—$CuO_2$-Y-$CuO_2$-BaO-MnO—LCMO and YBCO—$CuO_2$-BaO-CuO-(La,Ca)O-MnO—LCMO, where $CuO_2$ and CuO denote layers with square-planar and linear copper-oxide networks in YBCO, respectively. The former sequence with $CuO_2$ bilayers at the interface was found to be dominant [6]. However, the point resolution of these and other [8,9] conventional HRTEM and high-angle annular dark-field (HAADF) STEM experiments was limited by lens aberrations. Spherical aberration ($C_S$)-corrected instruments now available enable ultrahigh-resolution phase contrast (HRTEM) and ultrahigh-resolution Z-contrast (HAADF-STEM) imaging of transition metal oxide interfaces [16]. In particular, it was recently shown by $C_S$-corrected

STEM that mirror interfaces (A-B and B-A) in an oxide multilayer (-A-B-A-B- ) need not be identical, owing to the kinetics of pulsed laser deposition (PLD) growth [17]. Using the $C_S$-correction technique, it is now straightforward to obtain high-contrast HRTEM images of light elements (including oxygen) independent of their local environment, such that accurate quantification is possible. HAADF-STEM offers weaker contrast especially in situations in which light elements are surrounded by heavy elements. We have therefore chosen the former method in order to determine the atomic stacking sequence at the YBCO-LCMO interface.

The divergent reports regarding the magnetic properties of YBCO-LCMO heterostructures and the observation of different atomic stacking variants at oxide interfaces imply that magnetization measurements and HRTEM imaging have to be carried out on identical heterostructure samples in order to obtain quantitatively precise structure-property relationships. This requirement has motivated the present study of the magnetization and atomic positions of a single heterostructure comprised of a 7 unit cell (u.c.) thick LCMO film sandwiched between two 50 nm-thick YBCO films (inset in Fig. 1). The heterostructure was designed to ensure that top and bottom interfaces could be captured by a single HRTEM image.

The YBCO-LCMO-YBCO trilayer was grown on the (100) surface of $SrTiO_3$ by conventional PLD with an oxygen background pressure of 0.5 mbar at a pyrometrically controlled temperature of 780 °C, as described previously [2, 8, 12]. Individual YBCO (LCMO) films grown under identical conditions show epitaxial growth and a superconducting $T_C$ of 90 K (a Curie temperature of 250 K), close to the respective bulk values [3]. Fig. 1 shows the results of magnetization measurements on the trilayer sandwich. In spite of an LCMO thickness of only 7 u.c., the data reveal a Curie temperature of 235 K, comparable to that of bulk $La_{2/3}Ca_{1/3}MnO_3$ (Fig. 1a). The saturation magnetization $M_S = 230$ emu/cm$^3$ (Fig. 1b) is somewhat reduced with respect to the bulk value of 400 emu/cm$^3$ [18]. Possible origins of this reduction include a compensating ferromagnetic moment of YBCO and canting of the interfacial Mn spins due to superexchange across the interface [12], a modest charge transfer across the interface which pushes the interfacial LCMO layers closer to a charge-ordered antiferromagnetic state [13], and enhanced fluctuations due to the reduced dimensionality. We note that the $M_S$ we found is almost one order of magnitude higher than that of YBCO-LCMO superlattices with comparable LCMO thicknesses reported previously [5, 12].

Cross-sectional TEM specimens were prepared using standard techniques including grinding, dimpling, polishing, and ion milling [19]. During the preparation, great care was taken to avoid any contact of the specimen with water. Therefore, ethanol was used when grinding, dimpling, and polishing. The ion milling was done in a Fischione Ion Mill 1010 machine under 4.0 keV and 6.0 mA at an inclination angle of 10 degrees. The TEM experiments were carried out using a 300kV TEM (FEI Titan 80-300) with a field-emission gun and a spherical aberration corrector for the objective lens. The point-to-point resolution of the microscope is 0.78Å at 300 kV [20]. In our experiments, the HRTEM images were taken on a 1024 × 1024 pixel CCD camera at a magnification of 1.0 Mx at a sampling rate of 0.018 nm/ pixel. The alignment of the $C_S$-corrector was done using the CEOS software [21] based on aberration measurements deduced from Zemlin tableaus [22]. After iterative corrections using the amorphous area at the edge of the specimen, we achieved sufficiently small aberration coefficients. HRTEM images were recorded with small positive $C_S$ (< 1 μm) and small higher-order aberration coefficients.

$C_S$-corrected high-resolution images of the YBCO-LCMO-YBCO trilayer clearly show columns of all constituent atoms including the light oxygen atom columns (Fig. 2a.) Combined with image simulations of the individual YBCO and LCMO layers based on the multislice method [23], the TEM specimen thickness was determined to be 9.0 nm. All atom columns can be unambiguously identified (Fig. 2b) by overlapping the simulated images of the unit cells of YBCO and LCMO on the experimental image (see Fig. 2a, where two clippings of simulated images are superposed on the experimental image). Under our experimental conditions, the Y atom columns in the YBCO layer appear dark, the La atom columns in the LCMO layer are moderately bright, and the O atom columns are the brightest. The relative intensities and distances of all atom columns in the simulated image are consistent with the experimental ones, except the separation of O atoms around Y in a $CuO_2$-Y-$CuO_2$ bilayer unit which is smaller than observed. By an extensive set of simulations we established that this discrepancy originates in a failure of the simulation software to correctly account for the electron scattering in the relatively wide channel between Y atoms if the specimen thickness is relatively large.

In order to determine the atomic structures at the YBCO-LCMO and LCMO-YBCO interfaces, profiles of the image intensity across the interface were numerically averaged over a rectangular area of length 6.0 nm along the interface (marked by the framed box in Figs. 2a). By analyzing the intensity variations across the interface (Fig. 2c) and comparing them with those obtained from bulk areas away from the interfaces, it is clearly seen that two different interface configurations are present, which are highlighted by red fonts in Figs. 2c. The atomic stacking sequences at the top and bottom interfaces are YBCO—BaO-$CuO_2$-(La,Ca)O-MnO—LCMO and LCMO—(La,Ca)O-MnO-BaO-CuO-BaO-$CuO_2$-Y-$CuO_2$—YBCO, respectively. A corresponding atomic model is shown in Fig. 2b, where the individual atom columns are readily identified. It is further confirmed that the thickness of the LCMO layer is 7 u.c., i.e. about 2.7 nm. Other possible stacking sequences are inconsistent with the experimental intensity pattern.

To consolidate the observation based on one single $C_S$-corrected HRTEM image, a focus series of 15 images was recorded for exit-plane wave (EPW) reconstruction (TrueImage Focal-Series Reconstruction Package by FEI Company). The reconstructed amplitude and phase images of the EPW are shown in Fig. 3. Although the contrast in the reconstructed images is not as quite sharp as in the single-shot $C_S$-corrected HRTEM image (probably due to slight thickness variations and amorphization of the YBCO structure induced by progressive electron beam damage), two distinct interfaces can be indentified. In the reconstructed phase image strong dark dots in the YBCO layer denote $CuO_2$ columns, and Y atom columns appear as weaker dark dots. In the reconstructed amplitude image both columns appear as bright dots. The CuO atom contrast, however, is relatively low. (La,Ca)O atom columns in LCMO are small bright dots (hard to discern on the left side of the LCMO layer due to the small thickness variation), and O atom columns are weaker bright big dots in both reconstructed phase and amplitude images. Two simulated phase images using the interface atomic models on the right side are inserted, which agree with the reconstructed phase image (except for differences in the $CuO_2$-Y-$CuO_2$ unit already noted above). It can be seen that the interface atomic model deduced from reconstructed images yields the same configuration as Fig. 2a, which further supports the result obtained from the analysis of the single $C_S$-corrected HRTEM image above.

In summary, we have shown that the interfaces in our YBCO-LCMO-YBCO trilayer are epitaxial, atomically smooth, and uniform across the entire specimen segment captured by our TEM experiments. We have also demonstrated that the atomic stacking sequence at the interface of the LCMO layer grown on YBCO is different from that of the YBCO layer grown on LCMO, as previously reported for other oxide heterostructures [17]. Surprisingly, both sequences are different from the most common configuration previously identified in YBCO-LCMO superlattices [6]. This may imply that the interfacial stacking depends sensitively on the growth conditions. We note, however, that the atomic configurations we have identified on the YBCO side of the interfaces have recently also been observed by scanning tunnelling microscopy on a free YBCO surface [24].

The observed propensity to form different stacking sequences at interfaces calls for systematic experiments on the physical properties and atomic structure of identical YBCO-LCMO heterostructures. We have laid the foundation for such a study by determining the magnetic properties of a PLD-grown heterostructure with a 7 u.c. thick LCMO layer and atomically sharp interfaces. We found that the saturation magnetization $M_S$ is within a factor of two of that in bulk LCMO, and an order of magnitude larger than that of YBCO-LCMO heterostructures with comparable LCMO thicknesses grown by sputter deposition [5, 12]. Given the modest reduction of $M_S$ compared to bulk LCMO observed in our system, this large discrepancy seems unlikely to originate in differences in interfacial stacking patterns alone. Future experimental studies should therefore address possible extrinsic origins of the strongly reduced magnetization in these structures, as well as the chemical composition of heterostructures grown by different methods.

**References**

1. For reviews, see A. Buzdin, Rev. Mod. Phys. **77**, 935 (2005); I.F. Lyuksyutov and V. L. Pokrovsky, Adv. Phys. **54**, 67 (2005).
2. H.-U. Habermeier, G. Cristiani, R.K. Kremer, O. Lebedev, and G. Van Tendeloo, Physica C **364**, 298 (2001).
3. S. Soltan, J. Albrecht, and H.-U. Habermeier, Phys. Rev. B. **70**, 144517 (2004).

4. Z. Sefrioui, M. Varela, V. Peña, D. Arias, C. León, J. Santamaría, J.E. Villegas, and J.L. Martínez, Appl. Phys. Lett. **81**, 4568 (2002).
5. Z. Sefrioui, D. Arias, V. Peña, J.E. Villegas, M. Varela, P. Prieto, C. León, J.L. Martínez , and J. Santamaria, Phys. Rev. B **67**, 214511 (2003).
6. M. Varela, A.R. Lupini, S.J. Pennycook, Z. Sefrioui, and J. Santamaria, Solid-State Elec. **47**, 2245 (2003).
7. P. Przyslupski, I. Komissarov, W. Paszkowicz, P. Dluzewski, R. Minikayev, and M. Sawicki, Phys. Rev. B **69**, 134428 (2004).
8. T. Holden, H.-U. Habermeier, G. Cristiani, A. Golnik, A. Boris, A. Pimenov, J. Humlicek, O.I. Lebedev, G. Van Tendeloo, B. Keimer, and C. Bernhard, Phys. Rev. B **69**, 64505 (2004).
9. Z. Q. Yang, R. Hendrikx, J. Aarts, Y. Qin, and H.W. Zandbergen, Phys. Rev. B **67**, 024408 (2003).
10. V. Peña, Z. Sefrioui, D. Arias, C. Leon, J. Santamaria, J.L. Martinez, S.G.E. te Velthuis, and A. Hoffmann, Phys. Rev. Lett. **94**, 057002 (2005).
11. J. Stahn, J. Chakhalian, Ch. Niedermayer, J. Hoppler, T. Gutberlet, J. Voigt, F. Treubel, H.-U. Habermeier, G. Cristiani, B. Keimer, and C. Bernhard, Phys. Rev. B **71**, 140509(R) (2005).
12. A. Hoffmann, S. Velthuis, Z. Sefrioui, J. Santamaria, M.R. Fitzsimmons, S. Park, and M. Varela, Phys. Rev. B **72**, 140407(R) (2005).
13. J. Chakhalian, J.W. Freeland, G. Srajer, J. Strempfer, G. Khaliullin, J.C. Cezar, T. Charlton, R. Dalgliesh, C. Bernhard, G. Cristiani, H.-U. Habermeier, and B. Keimer, Nature Phys. **2**, 244 (2006).
14. J. Chakhalian, J.W. Freeland, H.-U. Habermeier , G. Cristiani, G. Khaliullin, M. van Veenendaal, B. Keimer, Science **318**, 1114 (2007).
15. J.W. Freeland, J. Chakhalian, H.-U. Habermeier, G. Cristiani, and B. Keimer, Appl. Phys. Lett. **90**, 242502 (2007).
16. For reviews, see D. A. Muller, L. Fitting Kourkoutis, M. Murfitt , J. H. Song, H. Y. Hwang, J. Silcox, N. Dellby, and O. L. Krivanek, Science **319**, 1073 (2008); K. W. Urban, Science **321**, 506 (2008).
17. L. Fitting Kourkoutis, D.A. Muller, Y. Hotta, and H.Y. Hwang, Appl. Phys. Lett. **91**, 163101 (2007).
18. H.L. Ju and H. Sohn, J. Magn. Magn. Mat. **167,** 200 (1997).
19. A. Strecker, U. Salzberger, and J. Mayer, Prakt. Metallogr. **30**, 482 (1993).
20. For a technical description, see: http://www.uni-ulm.de/einrichtungen/hrem.
21. S. Uhlemann and M. Haider, Ultramicroscopy **72**, 109 (1998).
22. F. Zemlin, K.Weiss, P. Schiske, W. Kunath, and K.-H. Herrmann, Ultramicroscopy **3**, 49 (1978).
23. A. Chuvilin and U. Kaiser, Ultramicroscopy **104**, 73 (2005).
24. G. Urbanik, T. Hanke, C. Hess, B. Büchner, A. Ciszewski, V. Hinkov, C.T. Lin, and B. Keimer, Eur. Phys. J. B **69**, 483 (2009).

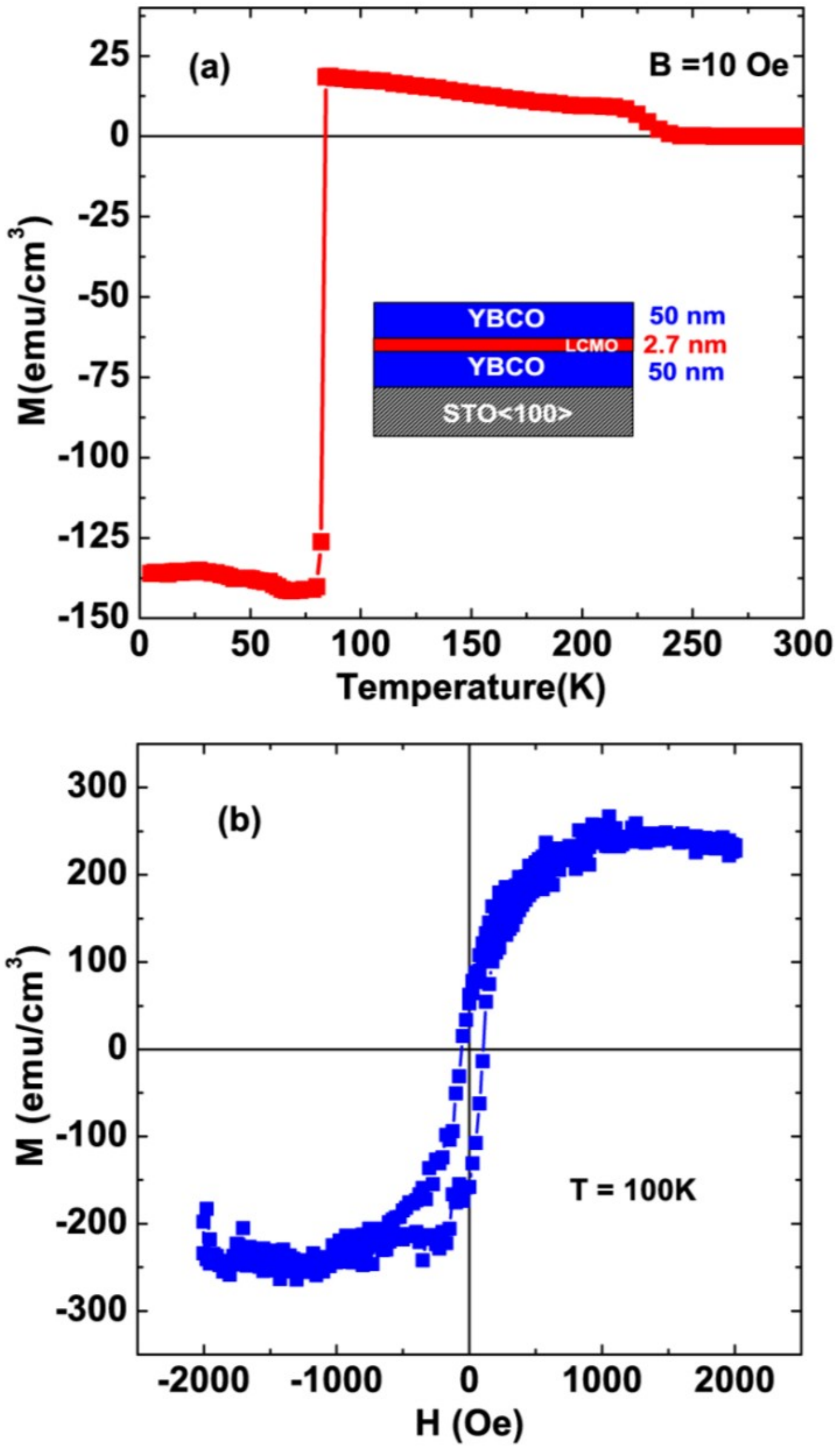

**Fig. 1**. (a) Field-cooled magnetization of the YBCO-LCMO-YBCO trilayer sketched in the inset in a magnetic field B = 10 Oe applied parallel to the layers. (b) Hysteresis loop at T = 100 K, above the superconducting transition of YBCO.

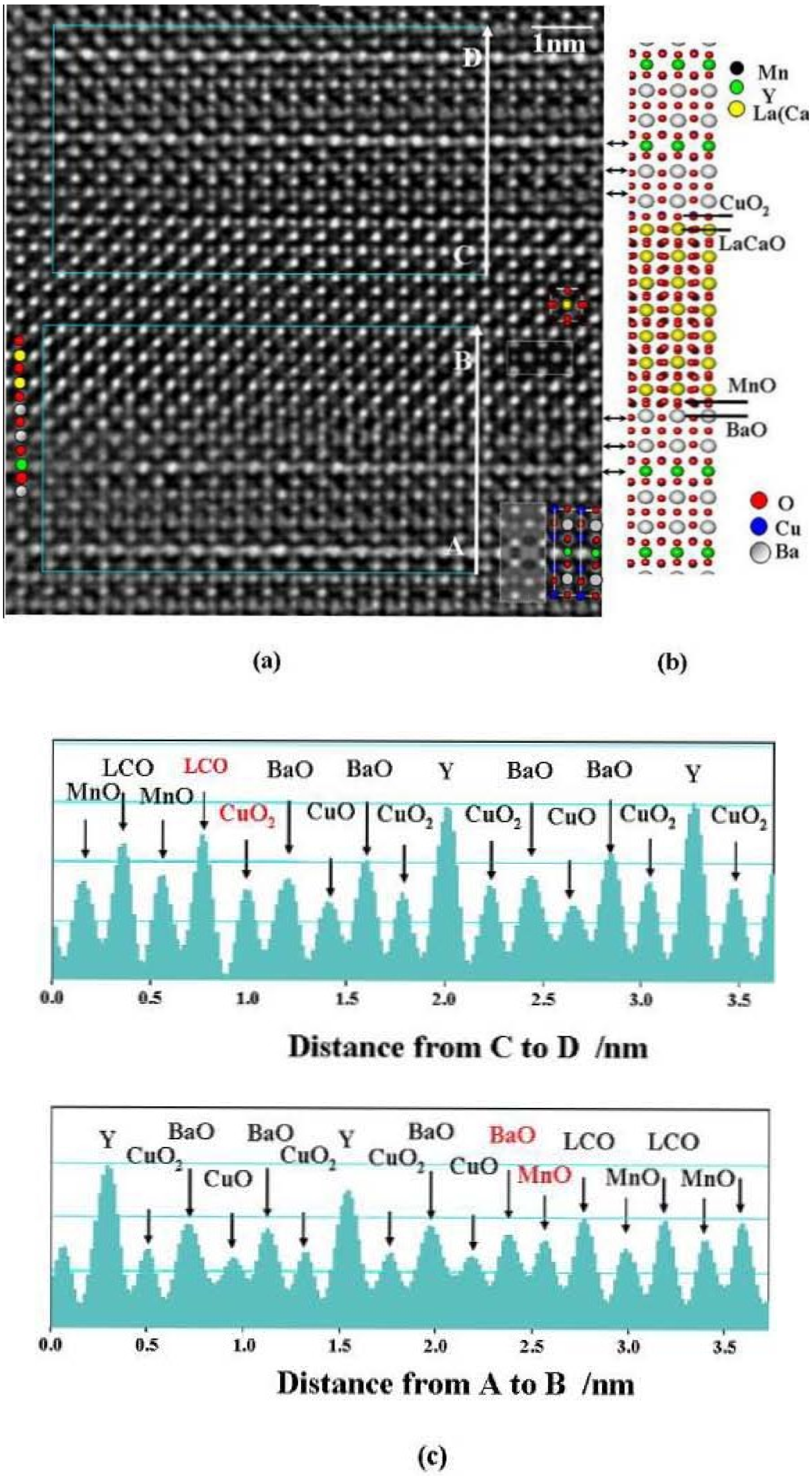

**Fig. 2**. (a) $C_S$-corrected cross-sectional HRTEM image of the central area of the YBCO-LCMO-YBCO trilayer, comprising the entire 7 u.c. thick LCMO layer and both interfaces. (b) Model of the atomic positions in the trilayer. Two clippings of the simulated images of bulk YBCO and LCMO based on the atomic model are superimposed on the experimental image (labelled by dotted lines at the top left corner and center left), which clearly show the respective atom column under the experimental conditions. (c) Line profiles of the image intensity [from A to B and from C to D in panel (a)] averaged over a rectangular area of length 6.0 nm. The interfacial planes are highlighted by red fonts. By comparing the intensity variations with those in bulk areas away from interfaces, two distinct interface configurations were identified and marked by black lines in panel (b).

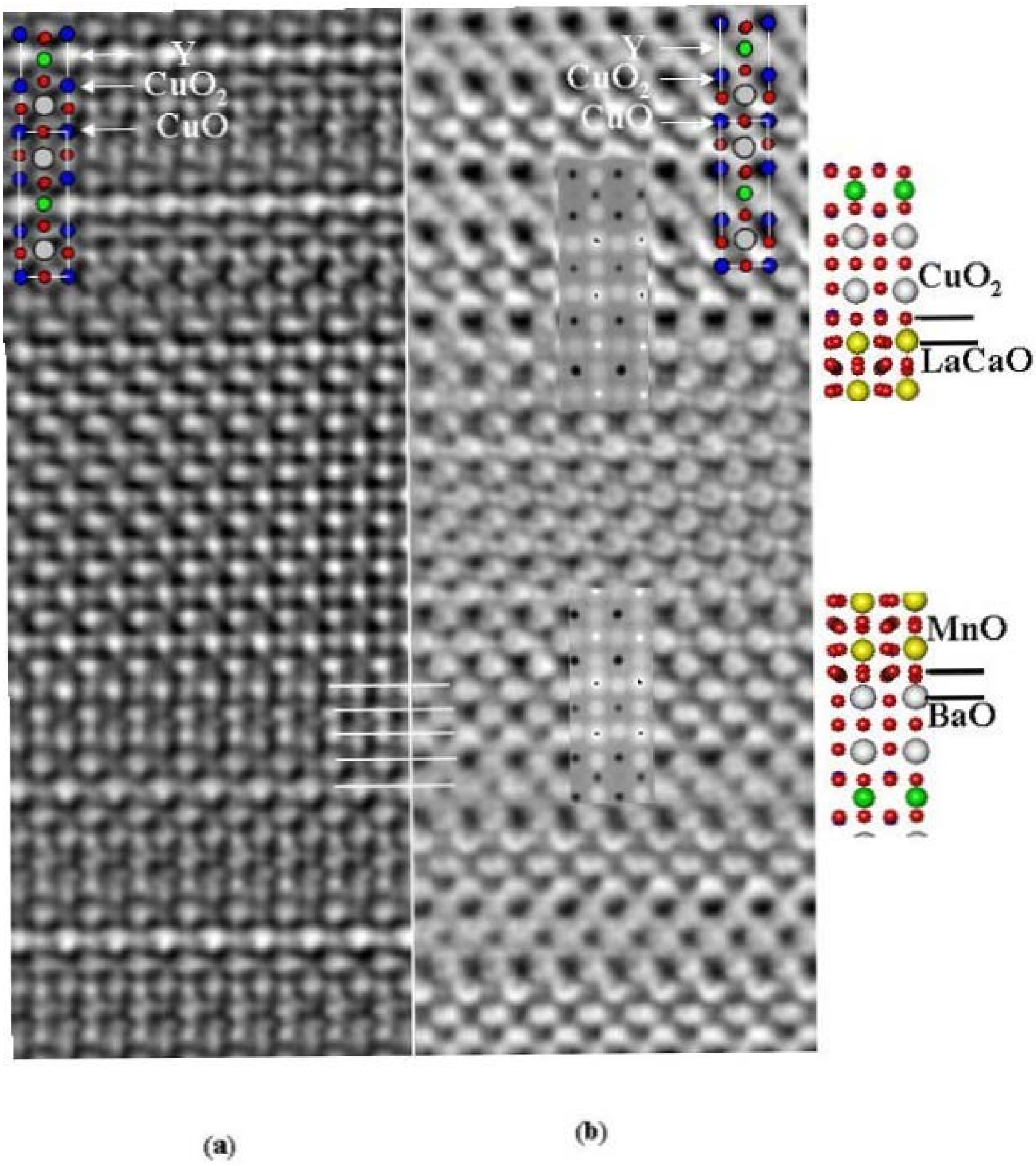

**Fig. 3**. (a) Amplitude and (b) phase of the reconstructed exit plane wave (EPW) of the YBCO-LCMO-YBCO trilayer. The YBCO unit atomic model is displayed as an inset in panels (a) and (b). The insets in panel (b) also show the results of the simulation for the upper and lower interface configurations shown on the right hand side, for a specimen thickness of 8.6 nm. White lines are guides-to-the-eye, indicating the corresponding planes in the reconstructed amplitude and phase images.